\documentclass[aps,prd,nofootinbib,twocolumn,superscriptaddress,floatfix]{revtex4}
\usepackage{mathrsfs}
\usepackage{latexsym}
\usepackage{amsmath}
\usepackage{amssymb}
\usepackage{graphicx}
\usepackage{longtable}
\usepackage{multirow}
\usepackage{bbm}
\usepackage{epsfig}
\usepackage[usenames]{color}
\usepackage[colorlinks=true,citecolor=blue,linkcolor=blue]{hyperref}
\allowdisplaybreaks

\begin{document}



\title{Speed of sound and liquid-gas phase transition in nuclear matter}

\author{Wei-bo He} 
\affiliation{ MOE Key Laboratory for Non-equilibrium Synthesis and Modulation of Condensed Matter, School of Physics, Xi’an Jiaotong University, Xi’an 710049, China}
\affiliation{ School of Physics, Peking University, Beijing, 100871, China}

\author{Guo-yun Shao}   
\email[Corresponding author: ]{gyshao@mail.xjtu.edu.cn} 
\affiliation{ MOE Key Laboratory for Non-equilibrium Synthesis and Modulation of Condensed Matter, School of Physics, Xi’an Jiaotong University, Xi’an 710049, China}


\author{Chong-long Xie}
\affiliation{ MOE Key Laboratory for Non-equilibrium Synthesis and Modulation of Condensed Matter, School of Physics, Xi’an Jiaotong University, Xi’an 710049, China}


\begin{abstract}
We investigate the speed of sound in nuclear matter at finite temperature and density~(chemical potential) in the nonlinear Walecka model. The numerical results suggest that the behaviors of sound speed are closely related to the the nuclear liquid-gas (LG) phase transition and the associated spinodal structure. 
The adiabatic sound speed is nonzero at the critical endpoint (CEP) in the mean field approximation. We further derive the boundary of vanishing sound velocity in the temperature-density phase diagram, and point out the region where the sound wave equation is broken. The distinction between the speed of sound in nuclear matter and that in quark matter contains important information about the equation of state of strongly interacting matter at intermediate and high density. We also formulate the relations between differently defined speed of sound using the fundamental thermodynamic relations.  
\end{abstract}

\maketitle
\section{introduction}

Exploring the equation of state~(EOS) of strongly interacting matter is an important topic in both the theoretical and experimental nuclear physics~\cite{Gupta11,Bazavov17,Borsanyi20,Fukushima04,Ratti06,Costa10, Sasaki12, Ferreira14, Schaefer10, Skokov11, Qin11, Gao16, Fischer14, Fu20,Stephanov,Shao2018,Shao2019,Luo2014, Luo2016, Luo2017,Adam21,Song11, Song112, Deb16,Leonhardt2020}.
During the space-time evolution of the newly formed matter in heavy-ion collisions, the speed of sound is one of the crucial  physical quantities to describe the variation of EOS. The dependence of speed of sound on environment  (temperature, density, chemical potential, etc.) carries important information in describing the evolution of the fireball and final observables.
Recently, the studies in \cite{Gardim20,Sahu21,Biswas20} show that the speed of sound  as a function of charged particle multiplicity $\langle d N_{ch}/d\eta \rangle$  can be extracted from heavy-ion collision data. In \cite{Sorensen21} the authors try to build a connection between the sound velocity and the baryon number cumulants to study the quantum chromodynamics~(QCD) phase structure. 

It is also an interesting topic to study the speed of sound during the QCD phase transition in the early universe by observing the induced gravitational wave. Although the propagation of gravitational wave is not sensitive to sound velocity, the value of sound velocity affects the dynamics of primordial density perturbations, and the induced gravitational waves by their horizon reentry can then be an indirect probe on both the EOS and sound velocity, which can provide useful knowledge of the evolution in the era of QCD phase transition~\cite{Abe21}. 

Besides in heavy-ion collision experiments and early universe, the speed of sound in neutron star matter also receives a lot of attention~(e.g.,\cite{Reed20, Kanakis20,Han20}). The density dependent behavior of sound velocity influences the mass-radius relation, the tidal deformability, and provides a sensitive probe of the EOS of neutron star matter and the hadron-quark phase transition in the dense core. To obtain a massive neutron star, some studies show that it is essential for neutron star matter to have a density range where the EOS is very stiff and the corresponding squared speed of sound is significantly larger than $1/3$~\cite{Tews18,Greif19,Forbes19,Drischler20,Essick20,Han19,Kojo,Altiparmak2022, Moustakidis2017, Zhang2019, Reed2020, Kanakis2020, Huang2022}. In addition, the study in \cite{Jaikumar21} indicates that the speed of sound is crucial for the gravitational wave frequencies induced by the $g$-mode oscillation of a neutron star. 

As an important quantity in describing the evolution of strongly interacting matter, the  speed of sound in QCD matter has been calculated, e.g.,  in LQCD~\cite{Aoki06, Borsanyi14, Bazavov14, Philipsen13, Borsanyi20},  (P)NJL model~\cite{Motta18,Ghosh06,Marty13,Deb16,Saha18}, quark-meson coupling model~\cite{Schaefer10,Abhishek18},  hadron resonance gas (HRG) model~\cite{Venugopalan92,Bluhm14}, field correlator method~(FCM)~\cite{Khaidukov18, Khaidukov19} and quasiparticle model~\cite{Mykhaylova21}.
In most studies, the main focus is put on the region of high temperature and  vanishing or small chemical potential. 

In Ref.~\cite{he2022}, we give an intensive study on the speed of sound of QCD matter in the full phase diagram. The numerical results suggest that the dependence of sound speed on temperature, density and chemical potential are closely related to the QCD phase structure.  In particular, the value of adiabatic sound speed is not zero at the CEP in the mean-field approximation.
In the region of chiral crossover region, the speed of sound  increases quickly with the rising temperature. The value of squared sound speed approaches to 1/3 after the chiral restoration at high temperature. Some new features of speed of sound are also discovered, for example, the hierarchy phenomenon of sound velocity for $u(d)$ and $s$ quark at low temperature with the increasing  chemical potential.
In addition to the adiabatic sound velocity, the behaviors of speed of sound under other conditions are also discussed. 

The nuclear LG phase transition was discovered in experiments many years ago, and the experimental phenomenon was explained with the spinodal decomposition mechanism~\cite{Chomaz2004,Pochodzalla95,Borderie01,Botvina95,Agostino99,Srivastava02,Elliott02,Ma1999,Deng2022}. Recently, there are some interesting investigations on  baryon number fluctuations as induced in the phase diagram
due to the presence of the nuclear LG phase transition~\cite{Vovchenko17, Shao20202}.  Since the nuclear LG transition and quark chiral first-order phase transition belong to the same universality class, it is interesting to study whether the speed of sounds in
quark matter and in nuclear matter behave in a similar way. Moreover, the study on the parameter dependence of the speed of sound in nuclear matter is crucial for investigating the transport phenomenon of hadronic matter~\cite{Deb16}. 
In this study, we will detailedly explore the behavior of sound speed in nuclear matter and compare it with that in quark matter. It is expected that the correlation between the behaviors of speed of sound and nuclear LG phase transition can be revealed.

The paper is organized as follows. In Sec.~II, we introduce briefly the nonlinear Walecka model  and  the formulae of speed of sound under different definitions. In Sec.~III, we present the numerical results of squared sound speed and discuss the relations with the nuclear LG transition. A summary is finally given in Sec. IV.

\section{the nonlinear Walecka model and speed of sound }

The Lagrangian density for the nucleons-meson system in the nonlinear Walecka model\cite{Glendenning1997} is 
\begin{eqnarray}\label{lagrangian}
\cal{L}\!&\!=\!&\sum_N\bar{\psi}_N\!\big[i\gamma_{\mu}\partial^{\mu}\!-\!(\!m_N
          \!-\! g_{\sigma }\sigma)\!
                 \! -\!g_{\omega }\gamma_{\mu}\omega^{\mu}  \big]\!\psi_N       \nonumber\\
         & &    +\frac{1}{2}\left(\partial_{\mu}\sigma\partial^
{\mu}\sigma-m_{\sigma}^{2}\sigma^{2}\right)\!-\! \frac{1}{3} bm_N\,(g_{\sigma} \sigma)^3-\frac{1}{4} c\,
(g_{\sigma} \sigma)^4
                    \nonumber\\
       & &+\frac{1}{2}m^{2}_{\omega} \omega_{\mu}\omega^{\mu}
          -\frac{1}{4}\omega_{\mu\nu}\omega^{\mu\nu}  
 \end{eqnarray}
 where $
\omega_{\mu\nu}= \partial_\mu \omega_\nu - \partial_\nu
\omega_\mu$. The interactions between
nucleons are mediated by $\sigma,\,\omega$ mesons.  The $ \mathrm{\rho}$ meson is not included  since in this work we only consider the behavior of speed of sound  in symmetric nuclear matter.  
The model parameters, $g_\sigma, g_\omega, b$ and $ c$, are fixed in the mean-field approximation with the compression modulus $K=240\,$MeV, the symmetric energy  $a_{sym}=31.3\,$MeV, 
the effective nucleon mass $m^*_N=m_N- g_\sigma \sigma=0.75m_N$~($m_N$ is the nucleon mass in vacuum) and the
 binding energy $B/A=-16.0\,$MeV at nuclear saturation density with $\rho_0=0.16\, fm^{-3}$.

 The thermodynamic  potential derived under the mean-field approximation is
\begin{eqnarray}
 \Omega\!&\!=&\!-\!\beta^{-1} \sum_{N} 2\! \int \frac{d^{3} \boldsymbol{k}}{(2 \pi)^{3}}\!\bigg[\ln \!\left(1+e^{-\beta\left(E_{N}^{*}(k)-\mu_{N}^{*}\!\right)}\right)\! \nonumber \\ 
 &&+\!\ln\! \left(1\!+\!e^{-\beta\left(E_{N}^{*}(k)+\mu_{N}^{*}\right)}\right)\!\bigg]\! +\!\frac{1}{2} m_{\sigma}^{2} \sigma^{2}\!+\!\frac{1}{3} b m_{N}\left(g_{\sigma} \sigma\right)^{3} \nonumber \\
 &&+\frac{1}{4} c\left(g_{\sigma} \sigma\right)^{4}-\frac{1}{2} m_{\omega}^{2} \omega_{}^{2},
  \end{eqnarray}
where $\beta=1/T$, $E_{N}^{*}=\sqrt{k^{2}+m_{N}^{*2}}$.  The effective chemical potential $\mu_{N}^{*}$ is defined as  $\mu_{N}^{*}=\mu_{N}-g_{\omega} \omega_{}$ for nucleons.

By minimizing the thermodynamical potential
\begin{equation}
\frac{\partial \Omega}{\partial \sigma}=\frac{\partial \Omega}{\partial \omega_{}}=0,
\end{equation}
the meson field equations are derived as
\begin{equation}\label{sigma}
g_{\sigma} \sigma\!=\!\left(\frac{g_{\sigma}}{m_{\sigma}}\right)^{2}\!\left[\rho_{p}^{s}+\rho_{n}^{s}-b m_{N}\left(g_{\sigma} \sigma\right)^{2}-c\left(g_{\sigma} \sigma\right)^{3}\right],
\end{equation}
\begin{equation}\label{omega}
g_{\omega} \omega=\left(\frac{g_{\omega}}{m_{\omega}}\right)^{2}\left(\rho_{p}+\rho_{n}\right).
\end{equation}

In Eqs.(\ref{sigma})-(\ref{omega}), the nucleon number density for proton or neutron is
\begin{equation}
\rho_N\!=\!2\! \int\! \frac{d^{3} \boldsymbol{k}}{(2 \pi)^{3}} \bigg[f\left(E_{N}^{*}(k)-\mu_{N}^{*}\right)\!-\!\bar{f}\left(E_{N}^{*}(k)+\mu_{N}^{*}\right) \bigg],
\end{equation}
and the scalar density
\begin{eqnarray}
\rho_{N}^{s}&=&2 \int \frac{d^{3} \boldsymbol{k}}{(2 \pi)^{3}} \frac{m_{N}^{*}}{E_{N}^{*}(k)}\bigg[f\left(E_{N}^{*}(k)-\mu_{N}^{*}\right)  \nonumber \\
& & \hspace{80pt}   +\bar{f}\left(E_{N}^{*}(k)+\mu_{N}^{*}\right)\bigg],
\end{eqnarray}
where $f(E_{N}^{*}(k)-\mu_{N}^{*})$ and $\bar{ f} (E_{N}^{*}(k)+\mu_{N}^{*})$ are the  fermion and antifermion distribution functions.

The pressure $p$ and energy density $\epsilon$ can be derived using the thermodynamic relations in the grand canonical ensemble as
\begin{equation}
\label{ }
p=-\Omega,\,\,\,\,\,\,\,\, \epsilon=-p+Ts+\sum \mu_N \rho_N,
\end{equation}
where $s$ is the entropy density. The general definition of speed of sound is
\begin{equation}
c^2_{X}=\left(\frac{\partial p}{\partial \epsilon}\right)_{X}.
\end{equation}
A specific constant quantity $X$ is required to describe the propagation of the compression wave through a medium. To indicate the different profiles of the EOS of nuclear matter, $X$ can be chosen as $s/\rho_B, s, \rho_B, T, \mu_B$.

For an ideal fluid, it evolves with a constant entropy density per baryon $s/\rho_B$ under the adiabatic evolution. This conclusion can be drawn in hydrodynamics with the conservation of energy and baryon number, therefore it is most meaningful to calculate the speed of sound along the isentropic curve
\begin{equation} 
\label{cs}
c_{s/{\rho_B}}^2=\bigg(\frac{\partial p}{\partial \epsilon}\bigg)_{s/{\rho_B}}.
\end{equation}
The dependence of $c_{s/{\rho_B}}^2$ on parameters, e.g., temperature, density and chemical potential, can indicate the variation of sound speed during the evolution and provide important knowledge of interaction, phase transition and  the EOS.

The definitions of sound velocity under other conditions are also taken in literature in dealing with different problems.
The speed of sound with constant baryon number density or entropy density are taken in describing the intermediate process of hydrodynamic evolution~\cite{Deb16,Albright16},
\begin{equation}
\label{ }
c_{\rho_B}^2=\bigg(\frac{\partial p}{\partial \epsilon}\bigg)_{\rho_B},
\end{equation}
and
\begin{equation}
\label{ }
c_s^2=\bigg(\frac{\partial p}{\partial \epsilon}\bigg)_s.
\end{equation}
For example, the space-time derivatives of temperature and  chemical potential are functions of  $c_{\rho_B}^2$ and $c_s^2$ with
\begin{equation}
\label{ }
\partial_0 \mu_B=-c_{s}^2  \mu_B\, \bold{\nabla}\cdot \bold{u},    
\end{equation}
and 
\begin{equation}
\label{ }
\partial_0 T=-c_{\rho_B}^2  T\, \bold{\nabla}\cdot \bold{u},    
\end{equation}
where $\bold u$ denotes the space component of four-velocity. 

It is also interesting to calculate the sound velocity with a fixed temperature or chemical potential with
\begin{equation}
\label{ }
c_T^2=\bigg(\frac{\partial p}{\partial \epsilon}\bigg)_T,   
\end{equation}
and
\begin{equation}
\label{ }
c_{\mu_B}^2=\bigg(\frac{\partial p}{\partial \epsilon}\bigg)_{\mu_B}.
\end{equation}
$c_T^2$ is widely used in calculating the sound speed in neutron star matter. In \cite{Sorensen21} the authors  make a
connection between the logarithmic derivative of $c_T^2$ with respect
to baryon density and baryon number fluctuations, which can be
measured in experiment.

In this study, we will explore the speed of sound in nuclear matter under different definitions in the full temperature-density and temperature-chemical potential  spaces. 
Since the general definitions given above can only be used to calculate the speed of sound along special trajectories, it is necessary to derive the corresponding formulae as functions of $T$ and $\mu_B$~($\rho_B$)  to perform the calculation for any given temperature and density~(chemical potential).

Using the fundamental thermodynamic relations we derive the sound speed formulae under different conditions in terms of $T$ and $\mu_B$ as
\begin{widetext}
\begin{equation}
c_{s/\rho_B}^{2}=\frac{s \rho_{B}\left(\frac{\partial s}{\partial \mu_{B}}\right)_{T}-s^{2}\left(\frac{\partial \rho_{B}}{\partial \mu_{B}}\right)_{T}-\rho_{B}^{2}\left(\frac{\partial s}{\partial T}\right)_{\mu_{B}}+s \rho_{B}\left(\frac{\partial \rho_{B}}{\partial T}\right)_{\mu_{B}}}{\left(s T+\mu_{B} \rho_{B}\right)\left[\left(\frac{\partial s}{\partial \mu_{B}}\right)_{T}\left(\frac{\partial \rho_{B}}{\partial T}\right)_{\mu_{B}}-\left(\frac{\partial s}{\partial T}\right)_{\mu_{B}}\left(\frac{\partial \rho_{B}}{\partial \mu_{B}}\right)_{T}\right]},
\end{equation}
\begin{equation}
c_{\rho_{B}}^{2}=\frac{s\left(\frac{\partial \rho_{B}}{\partial \mu_{B}}\right)_{T}-\rho_{B}\left(\frac{\partial \rho_{B}}{\partial T}\right)_{\mu_{B}}}{T\left[\left(\frac{\partial s}{\partial T}\right)_{\mu_{B}}\left(\frac{\partial \rho_{B}}{\partial \mu_{B}}\right)_{T}-\left(\frac{\partial s}{\partial \mu_{B}}\right)_{T}\left(\frac{\partial \rho_{B}}{\partial T}\right)_{\mu_{B}}\right]},
\end{equation}
\begin{equation}
c_{s}^{2}=\frac{s\left(\frac{\partial s}{\partial \mu_{B}}\right)_{T}-\rho_{B}\left(\frac{\partial s}{\partial T}\right)_{\mu_{B}}}{\mu_{B}\left[\left(\frac{\partial \rho_{B}}{\partial T}\right)_{\mu_{B}}\left(\frac{\partial s}{\partial \mu_{B}}\right)_{T}-\left(\frac{\partial s}{\partial T}\right)_{\mu_{B}}\left(\frac{\partial \rho_{B}}{\partial \mu_{B}}\right)_{T}\right]},
\end{equation}
\begin{equation}
c_{T}^{2}=\frac{\rho_{B}}{T\left(\frac{\partial s}{\partial \mu_{B}}\right)_{T}+\mu_{B}\left(\frac{\partial \rho_{B}}{\partial \mu_{B}}\right)_{T}},
\end{equation}
\begin{equation}
c_{\mu_{B}}^{2}=\frac{s}{T\left(\frac{\partial s}{\partial T}\right)_{\mu_{B}}+\mu_{B}\left(\frac{\partial \rho_{B}}{\partial T}\right)_{\mu_{B}}}.
\end{equation}
\end{widetext}

The  corresponding sound speed formulae derived in terms of $T$ and $\rho_B$ are
\begin{widetext}
\begin{equation}
c^2_{s/\rho_B}=\frac{s^{2}+\rho_{B}^{2}\left[\left(\frac{\partial \mu_{B}}{\partial \rho_{B}}\right)_{T}\left(\frac{\partial s}{\partial T}\right)_{\rho_{B}}-\left(\frac{\partial \mu_{B}}{\partial T}\right)_{\rho_{B}}\left(\frac{\partial s}{\partial \rho_{B}}\right)_{T}\right]+s \rho_{B}\left[\left(\frac{\partial \mu_{B}}{\partial T}\right)_{\rho_{B}}-\left(\frac{\partial s}{\partial \rho_{B}}\right)_{T}\right]}{\left(T s+\mu_{B} \rho_{B}\right)\left(\frac{\partial s}{\partial T}\right)_{\rho_{B}}},
\end{equation}
\begin{equation}
c_{\rho_{B}}^{2}=\frac{s+\rho_{B}\left(\frac{\partial \mu_{B}}{\partial T}\right)_{\rho_{B}}}{T\left(\frac{\partial s}{\partial T}\right)_{\rho_{B}}},
\end{equation}
\begin{equation}
c_{s}^{2}=\frac{\rho_{B}\left[\left(\frac{\partial s}{\partial T}\right)_{\rho_{B}}\left(\frac{\partial \mu_{B}}{\partial \rho_{B}}\right)_{T}-\left(\frac{\partial s}{\partial \rho_{B}}\right)_{T}\left(\frac{\partial \mu_{B}}{\partial T}\right)_{\rho_{B}}\right]-s\left(\frac{\partial s}{\partial \rho_{B}}\right)_{T}}{\mu_{B}\left(\frac{\partial s}{\partial T}\right)_{\rho_{B}}},
\end{equation}
\begin{equation}
c_{T}^{2}=\frac{\rho_{B}\left(\frac{\partial \mu_{B}}{\partial \rho_{B}}\right)_{T}}{T\left(\frac{\partial s}{\partial \rho_{B}}\right)_{T}+\mu_{B}},
\end{equation}
\begin{equation}
c_{\mu_{B}}^{2}=\frac{s\left(\frac{\partial \mu_{B}}{\partial \rho_{B}}\right)_{T}}{T\left[\left(\frac{\partial s}{\partial T}\right)_{\rho_{B}}\left(\frac{\partial \mu_{B}}{\partial \rho_{B}}\right)_{T}-\left(\frac{\partial \mu_{B}}{\partial T}\right)_{\rho_{B}}\left(\frac{\partial s}{\partial \rho_{B}}\right)_{T}\right]-\mu_{B}\left(\frac{\partial \mu_{B}}{\partial T}\right)_{\rho_{B}}}.
\end{equation}
\end{widetext}

\section{Numerical results and discussions }

In this section, we present the phase structure of symmetric nuclear matter and the numerical results of the speed of sound under different definitions at finite temperature and baryon density~(chemical potential), and discuss the relations between the speed of sound and nuclear LG phase transition. Symmetric nuclear matter is considered in this study to describe the essential  behavior of speed of sound.

\subsection{Nuclear liquid-gas phase transition and speed of sound  at constant $s/\rho_B$}

\begin{figure*}[htbp]
\begin{center}
\includegraphics[scale=0.4]{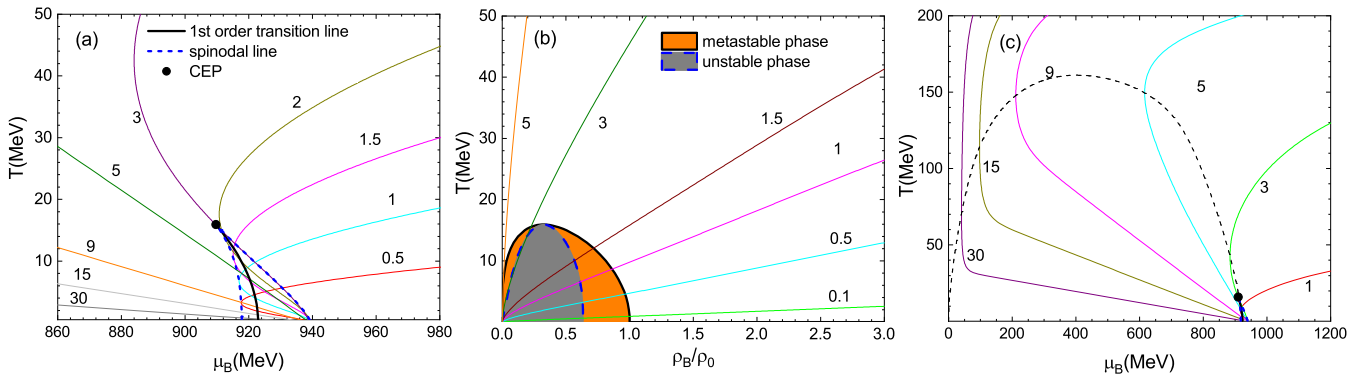}
\caption{  Phase diagram of nuclear liquid-gas transition and the isentropic curves. (a) spinodal structure of LG phase transition the  in $T-\mu_B$ plane (b) phase structure of LG phase transition in $T-\rho_B$ plane (c) isentropic trajectories in the full phase diagram, the black dashed line is derived with $(\frac{\partial \mu_B}{\partial T})_{s/\rho_B}=0$.}
\label{fig:1}   
\end{center}
\end{figure*}

Firstly, we demonstrate in Fig.~\ref{fig:1} the phase structure of nuclear LG phase transition and the isentropic curves for different values of $s/\rho_B$.
Fig.~\ref{fig:1} (a) and (b) show the details of nuclear LG phase transition and the corresponding spinodal structure. The black solid line is the first-order phase transition line, and the blue dashed line is the boundary of spinodal structure associated with the nuclear LG phase transition. Fig.~\ref{fig:1} (b) shows clearly that the two curves separate the phase diagram into the stable, metastable and unstable phases. The spinodal phase decomposition  plays a dominant role in the experimental exploration of the first-order nuclear LG transition\cite{Chomaz2004,Shao20202}. It has also inspired the anticipation to identify the first-order chiral phase transition in high-energy heavy-ion collisions through the spinodal phase separation~\cite{Mishustin1999, Randrup2004, Koch2005, Sasaki2007, Sasaki2008, Randrup2009, Steinheimer2012, Li2016, Steinheimer2016, Steinheimer2017}.

The first-order transition line and the spinodal line meet at the CEP. The isentropic curves in Fig.~\ref{fig:1} indicate the evolutionary trajectories of an ideal fluid under the adiabatic condition. Fig.~\ref{fig:1} (b) shows that for smaller $s/\rho_B$ the evolutionary trajectories pass through the stable, metastable and unstable phases. It is expected that the important information on the phase transition is carried by speed of sound. Fig.~\ref{fig:1} (c) shows the nuclear LG transition and isentropic trajectories in the full $T-\mu_B$ phase diagram. In Fig.~\ref{fig:1} (c), we also plot the boundary of $(\frac{\partial \mu_B}{\partial T})_{s/\rho_B}=0$~(black dashed curve), which is closely related to the behavior of sound speed at constant entropy density.

\begin{figure} [htbp]
\begin{minipage}{\columnwidth}
\centering
\includegraphics[scale=0.33]{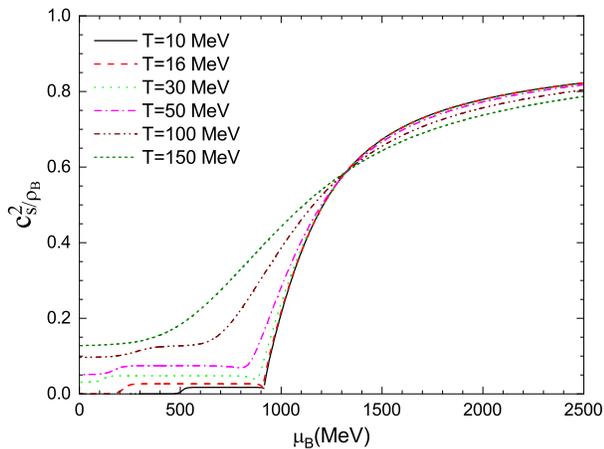}
\end{minipage}
\caption{ Squared sound speed $c^2_{s/\rho_B}$ as functions of chemical potential for several fixed temperatures.}
\label{fig:2}
\end{figure}

\begin{figure} [htbp]
\begin{minipage}{\columnwidth}
\centering
\includegraphics[scale=0.33]{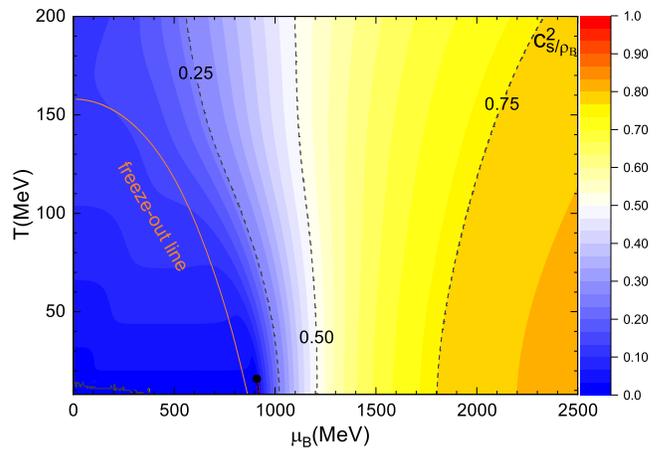}
\end{minipage}
\caption{ Contour map of $c^2_{s/\rho_B}$ in the  $T-\mu_B$ panel. }
\label{fig:3}
\end{figure}

We present the squared speed of sound $c^2_{s/\rho_B}$ as functions of baryon chemical potential for different temperatures in Fig.~\ref{fig:2} 
and the contour map in Fig.~\ref{fig:3}. Fig.~\ref{fig:2} shows that $c^2_{s/\rho_B}$ grows with the increase of chemical potential for each temperature. 
It also indicates that, for $\mu_B<1320\,$MeV, 
$c^2_{s/\rho_B}$ at a higher temperature is larger than that at a lower temperature, because the pressure and energy density is mainly driven by temperature for small chemical potential. For $\mu_B>1320\,$MeV, the opposite happens, which is mainly attributed to the density-driven effect with the decrease of dynamics mass of nucleon. For $T=10\,$MeV, a small jump of $c^2_{s/\rho_B}$ appears  on the boundary of the first-order phase transition.

\begin{figure} [htbp]
\begin{minipage}{\columnwidth}
\centering
\includegraphics[scale=0.33]{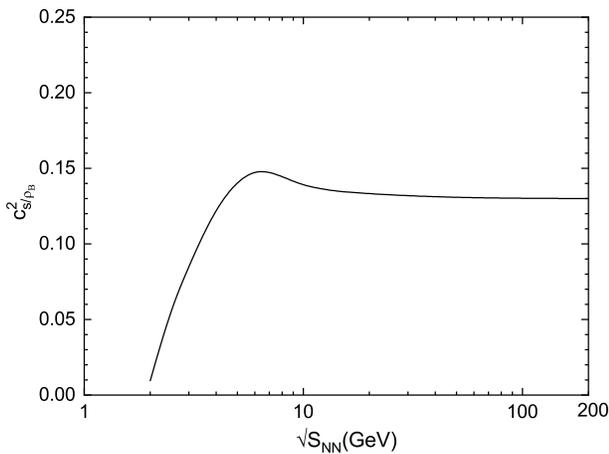}
\end{minipage}
\caption{ Values of $c^2_{s/\rho_B}$ along the chemical freeze-out line as a function of collision energy.}
\label{fig:4}
\end{figure}

The contour map in Fig.~\ref{fig:3} demonstrates the profiles of $c^2_{s/\rho_B}$ in the full $T-\mu_B$ phase diagram with the nuclear LG phase transition. In this figure we also plot the fitted chemical freeze-out line. The parameterized formula of the freeze-out line is taken from Ref.~\cite{Luo2017} with
\begin{equation}
\mu_{B}(\sqrt{s_{NN}})=\frac{1.477}{1+0.343 \sqrt{s_{NN}}}
\end{equation}
and
\begin{equation}
T\left(\mu_{B}\right)=a-b \mu_{B}^{2}-c \mu_{B}^{4},
\end{equation}
where $a=0.158$~GeV, $b=0.14~\text{GeV}^{-1}$, and  $c$ varies from 0.04 to 0.12. The chemical freeze-out curve in Fig.~\ref{fig:3} is plotted for the intermediate value of $c=0.08$. The value of $c^2_{s/\rho_B}$ along the freeze-out line is given in Fig.~\ref{fig:4}. It shows $c^2_{s/\rho_B}$ on the freeze-out line changes slightly for 
collision energy larger than 10 GeV. A small peak appears at  $\sqrt{s_{NN}}\simeq 6\,$GeV, and $c^2_{s/\rho_B}$ decreases with the decline of collision energy.

\begin{figure} [htbp]
\begin{minipage}{\columnwidth}
\centering
\includegraphics[scale=0.33]{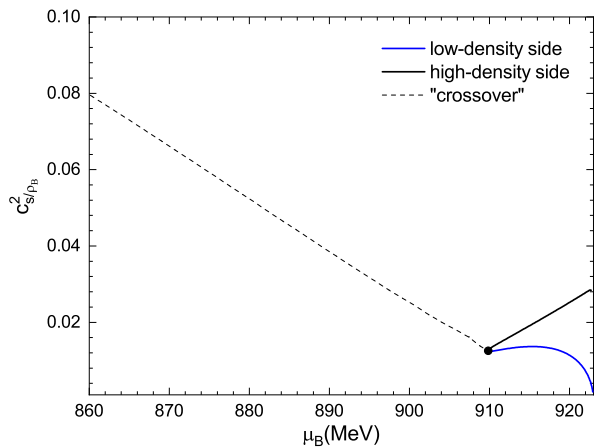}
\end{minipage}
\caption{ Values of $c^2_{s/\rho_B}$ on the boundary of first-order transition and the "crossover" line.}
\label{fig:5}
\end{figure}

We can also see that the chemical freeze-out line at low temperature is close to the nuclear LG phase transition. We display the value of $c^2_{s/\rho_B}$ on the boundary of the nuclear
LG phase transition in Fig.~\ref{fig:5}. The blue solid line is the value along the low-density side of the first-order phase transition and the black solid line 
is the result on the high-density side. The dash-dotted line is the ``crossover" line, which is derived with $\partial \sigma/ \partial \mu_B$  (or $\partial m^*_N/ \partial \mu_B$) taking the extremum for a given temperature. This is inspired by the description of quark chiral phase transition in quark models. The aim of plotting the ``crossover" line is to show that some thermodynamic properties are sensitive to this line. Such a description is also taken to discuss the net baryon number fluctuations induced by the nuclear LG phase transition~\cite{Shao20202}.

\begin{figure} [htbp]
\begin{minipage}{\columnwidth}
\centering
\includegraphics[scale=0.33]{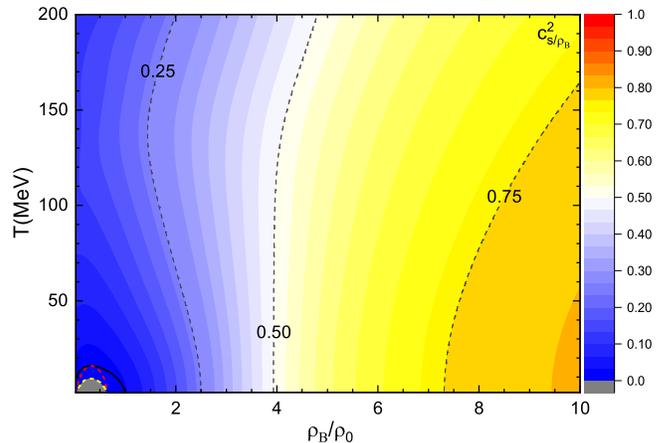}
\end{minipage}
\caption{ Contour map of $c^2_{s/\rho_B}$ in the  $T-\rho_B$ panel.}
\label{fig:6}
\end{figure}

Fig.~\ref{fig:5} indicates that the square of sound speed is not zero but a small value at the CEP. It is a feature of the mean field approximation.  A similar behavior exists
for the quark chiral phase transition in the PNJL model with the mean-field assumption~\cite{he2022} . Fig.~\ref{fig:5} also suggests that the value of $c^2_{s/\rho_B}$ is  small near the first-order transition region. Since only the stable phase is considered in the above discussion, it cannot give a complete demonstration of sound speed along the isentropic trajectories. Therefore, we further present in Fig.~\ref{fig:6}  the 
contour map of $c^2_{s/\rho_B}$ in the full $T-\rho_B$ panel, including all the stable, metastable and unstable phases.

In Fig.~\ref{fig:6}, besides the first-order phase transition line~(black solid line) and the spinodal line~(red dashed line), we also derive the boundary of vanishing sound speed, which is plotted with the yellow dashed line. At each point of this boundary, it fulfills the condition of $(\frac{\partial p}{\partial \epsilon})_{s/\rho_B}=0$. Inside this boundary~(grey area), the value of $c_{s/\rho_B}^2=(\frac{\partial p}{\partial \epsilon})_{s/\rho_B}$
 is negative. The sound wave equation is broken in this situation, and becomes a decay function. It means that a disturbance cannot be propagated in this region.

\begin{figure} [htbp]
\begin{minipage}{\columnwidth}
\centering
\includegraphics[scale=0.32]{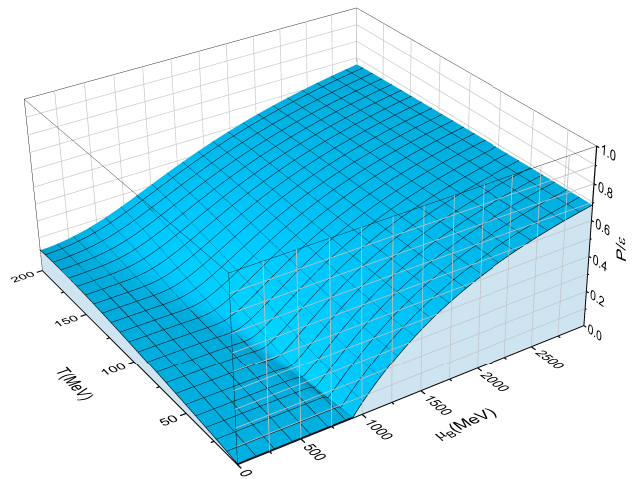}
\end{minipage}
\caption{ 3D map of $p/\epsilon$ as a function of temperature and chemical potential.}
\label{fig:7}
\end{figure}

From figures~\ref{fig:2},~\ref{fig:3} and~\ref{fig:6}, we can see that the square of sound speed is quite larger than $1/3$ at high density~(large chemical potential), about 0.8 being approached at very large chemical potential. 
However, the  value of $c^2_{s/\rho_B}$ in quark matter approaches to $1/3$ at high density~\cite{he2022}. 
To understand the variation of speed of sound in nuclear matter,  we plot the 3D map of $p/\epsilon$ as a function of  the temperature and chemical potential in Fig.~\ref{fig:7}. It indicates that the behavior of $p/\epsilon$ is responsible for the growing  speed of sound from low to high chemical potential.

\begin{figure} [htbp]
\begin{minipage}{\columnwidth}
\centering
\includegraphics[scale=0.33]{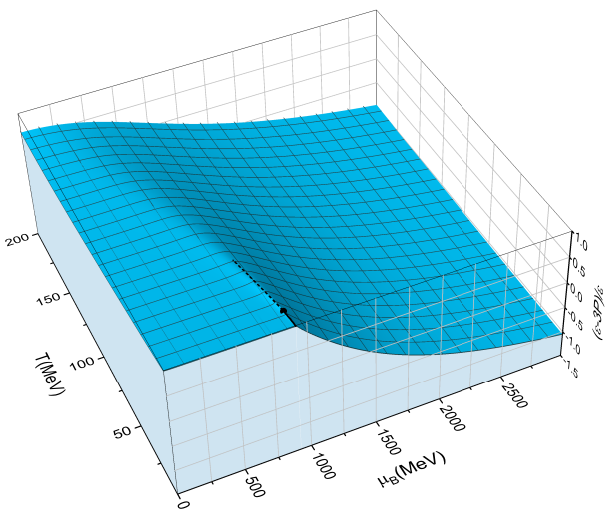}
\end{minipage}
\caption{ 3D map of $(\epsilon-3p)/\epsilon$ as a function of temperature and chemical potential. The black solid line is the nuclear LG phase transition line. The black dashed line is the  ``crossover'' line~\cite{Shao20202}}
\label{fig:8}
\end{figure}

The interaction measurement or trace anomaly is defined as $\epsilon-3p$, which can effectively describe the interaction in a thermal system.  We demonstrate the  
value of $(\epsilon-3p)/\epsilon$ of symmetric nuclear matter in Fig.~\ref{fig:8}.  
Comparing it with the phase transition line, we find the behavior of $(\epsilon-3p)/\epsilon$ is related to the nuclear phase transition.
The behavior of $(\epsilon-3p)/\epsilon$ reflects the variation of nucleon mass or the $\sigma$ meson field, i.e. the interaction between nucleons.  Moreover, we can see that $(\epsilon-3p)/\epsilon$ in nuclear matter does not approach zero at large chemical potential, quietly different from that of quark matter. 

The calculation in the Walecka model and PNJL model indicate that the speed of sound in nuclear matter is quite larger than that in quark matter at high density. Theoretically,  in the Walecka model, the
$\omega$ meson interaction is proportional to the baryon density,
leading to a steady increase in the speed of sound, with the limiting
value of 1 at $\rho_B \to \infty$. In the (P)NJL model, the value of
the sigma mean-field, and therefore of the corresponding interaction,
decreases with density. In view of this  the (P)NJL model looks like a gas of non-interacting relativistic
particles at $\rho_B \to \infty$. To the extent that the Walecka model
can be thought to describe nuclear matter and the (P)NJL model can
give one some insight of high-density quark matter.  If a phase transition from nuclear matter to quark matter takes place
with growing density, the  value of speed of sound  will have a peak at a certain density.  It is interesting that recent neutron star research suggests that the value of the speed of sound first increases sharply with $\rho_B$,
exceeding the conformal value of $1/3$, then falls again below $1/3$,
and finally approaches $1/3$ at infinity from below.  It is anticipated to give a further analysis on the speed of sound of neutron star matter with different phase transition mechanism and observation constraints.

\subsection {Speed of sound at constant $\rho_B$ or $s$}
\begin{figure} [htbp]
\begin{minipage}{\columnwidth}
\centering
\includegraphics[scale=0.33]{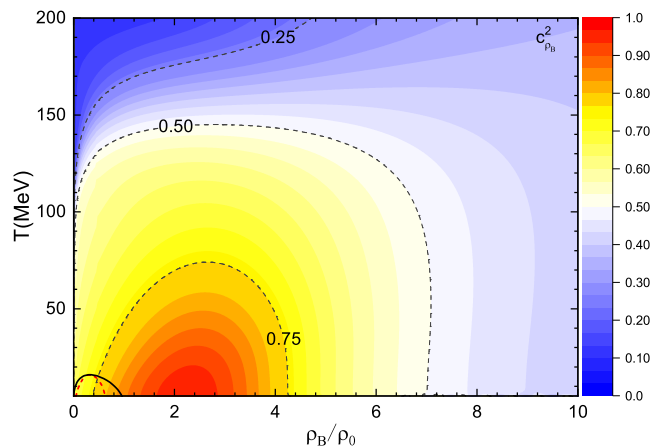}
\end{minipage}
\caption{ Contour map of $c^2_{\rho_B}$ in the  $T-\rho_B$ panel.}
\label{fig:9}
\end{figure}

\begin{figure} [htbp]
\begin{minipage}{\columnwidth}
\centering
\includegraphics[scale=0.33]{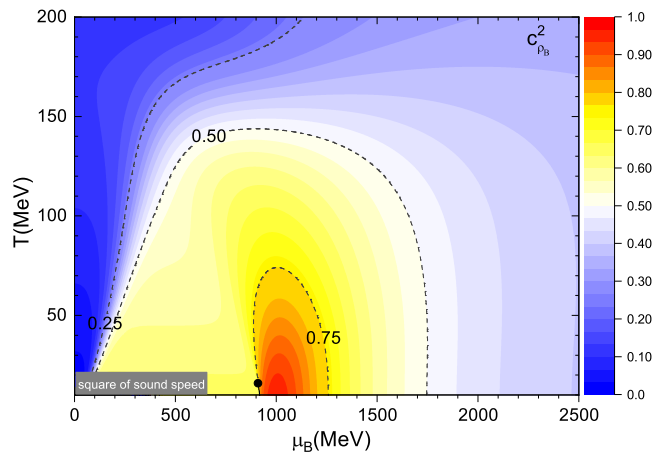}
\end{minipage}
\caption{ Contour map of $c^2_{\rho_B}$ in the  $T-\mu_B$ panel.}
\label{fig:10}
\end{figure}

We plot the contour map of the squared speed of sound $c^2_{\rho_B}$ in the $T-\rho_B$ panel in Fig.~\ref{fig:9} and $T-\mu_B$ panel in Fig.~\ref{fig:10}. The numerical results indicate that the value of $c^2_{\rho_B}$ lies in the range of $0 \sim 1$. For most temperatures, $c^2_{\rho_B}$ shows a nonmonotonous behavior with the increase of density or chemical potential. In particular, a peak-like structure exists at intermediate density~(chemical potential).


\begin{figure} [htbp]
\begin{minipage}{\columnwidth}
\centering
\includegraphics[scale=0.33]{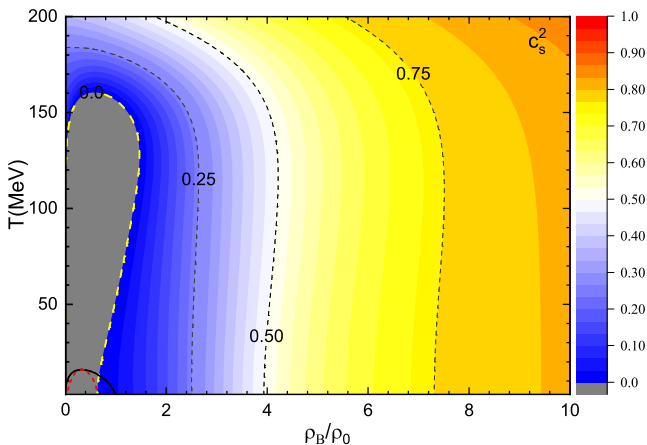}
\end{minipage}
\caption{ Contour map of $c^2_{s}$ in the  $T-\rho_B$ panel. The yellow dashed line is boundary where the value of $c^2_{s}$ vanishes. $c_s^2=({\partial p}/{\partial \epsilon})_{s}$ is negative inside the yellow dashed line.}
\label{fig:11}
\end{figure}

\begin{figure} [htbp]
\begin{minipage}{\columnwidth}
\centering
\includegraphics[scale=0.33]{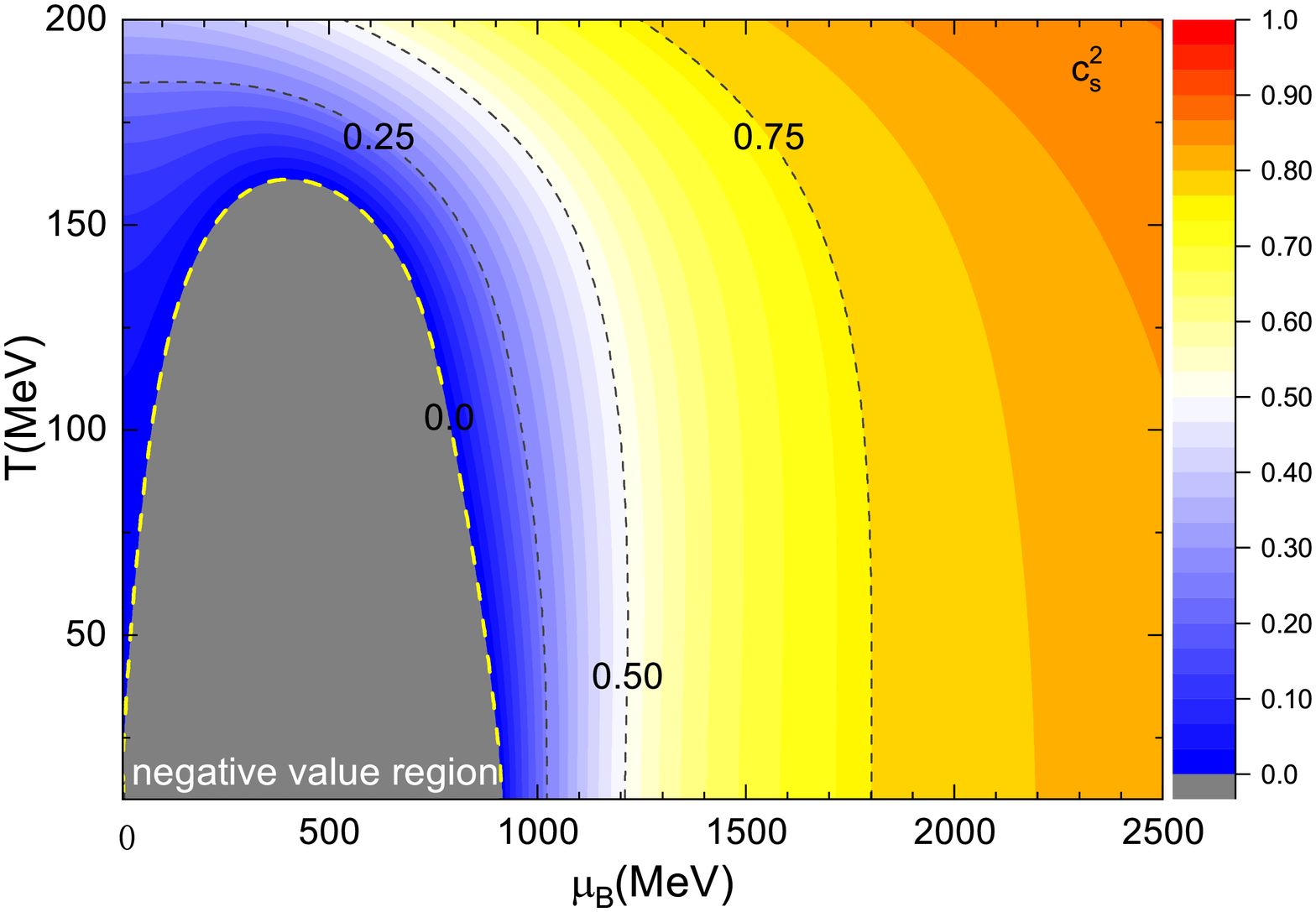}
\end{minipage}
\caption{ Contour map of $c^2_{s}$ in the  $T-\mu_B$ panel. The yellow dashed line is boundary where the value of $c^2_{s}$ vanishes.  $c_s^2=({\partial p}/{\partial \epsilon})_{s}$ is negative inside the yellow dashed line.}
\label{fig:12}
\end{figure}

We present the squared speed of sound $c^2_{s}$ at constant entropy density in Fig.~\ref{fig:11} and Fig.~\ref{fig:12}. The two figures indicate that the contour of $c^2_{s}$ has a relatively complicated structure. The value of  $c^2_{s}$ are divided into two parts by the yellow dashed curve. Outside the  curve,  the value of  $c^2_{s}$ is positive and smaller than 1. Inside the  curve~(gray area), $c^2_{s}$ is negative. The value of $c^2_{s}$ vanishes on the boundary given by the yellow dashed curve. Such a feature is possibly general for a first-order phase transition in a interacting system with temperature and density dependent fermion mass, and a similar structure for the speed of sound in quark matter is found in Ref.~\cite{he2022}.


As a matter of fact, the boundary~(yellow dashed line) in Fig.~\ref{fig:11} and Fig.~\ref{fig:12} is just the curve shown in Fig.~\ref{fig:1} (c). This can be proven with the thermodynamics formula  
\begin{equation}
\label{rhos}
\bigg(\frac{\partial \mu_B}{\partial T}\bigg)_{s/\rho_B}=\frac{\mu_B}{T}\frac{(\frac{\partial p}{\partial \epsilon})_{s}}{(\frac{\partial p}{\partial \epsilon})_{\rho_{B}}}=\frac{\mu_B}{T}\frac{c^2_s}{c^2_{\rho_B}}.
\end{equation}
Using Eqs.~(\ref{rhos}) we can obtain the boundary of vanishing $c^2_{s}$ by taking  $(\frac{\partial \mu_B}{\partial T})_{s/\rho_B}=0$. This is the physical condition used to derive the dashed curve in  Fig.~\ref{fig:1} (c). 
Moreover, when the condition $(\frac{\partial \mu_B}{\partial T})_{s/\rho_B}<0$ is fulfilled, one of the two physical quantities $c_s^2$ and $c_{\rho_B}^2$ takes a negative value. There indeed exists such a region~in the phase diagram, as shown in Fig.~\ref{fig:1} (c)~(inside the black dashed curve). Since $c_{\rho_B}^2$ is always positive, $c_s^2=({\partial p}/{\partial \epsilon})_{s}$ is negative in this situation. The blue areas in Fig.~\ref{fig:11} and Fig.~\ref{fig:12} indicate the negative region. 

Comparing the result with that of quark matter in the PNJL model~(Fig.~1 in Ref.~\cite{he2022} ), we find there are three regions in which  $(\frac{\partial \mu_B}{\partial T})_{s/\rho_B}<0$ for quark matter,
including  the lower left of the $T-\mu_B$ phase diagram and a small neighboring area of chiral-crossover transformation near the CEP, as well as a part of the region where the chiral symmetry of $u,d$ quarks is restored but the Polyakov loop is still confining.  Correspondingly, $c_s^2$ or $c_{\rho_B}^2$ is negative in the these regions. One can refer to Ref.~\cite{he2022} for details.

\subsection {Speed of sound at constant $T$ or $\mu_B$}
In the following, we study the speed of sound derived at constant $T$ or $\mu_B$. Recently, the density dependent $c^2_T$ has been discussed for neutron star matter under the equilibrium of weak interaction. The data from observations support a large value of $c^2_T$~(larger than 1/3) at a few times nuclear saturation density.  In \cite{Sorensen21} the authors also try to estimate $c^2_T$ at chemical freeze-out of quark-gluon plasma using the baryon number fluctuations of the beam energy scan experiments at RHIC.  We explore here the behavior of $c^2_T$ and $c^2_{\mu_B}$ in nuclear matter in the full $T-\rho_B$ and $T-\mu_B$ phase diagram. 

\begin{figure} [htbp]
\begin{minipage}{\columnwidth}
\centering
\includegraphics[scale=0.33]{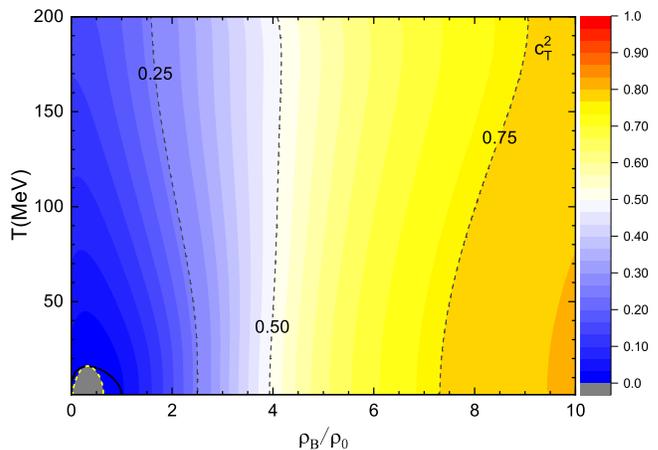}
\end{minipage}
\caption{ Contour map of $c^2_{T}$ in the  $T-\rho_B$ panel. The yellow dashed curve is the spinodal line and the boundary of $c^2_{T}=0$. $c^2_{T}=(\partial p / \partial \epsilon)_{T}$ is negative inside the boundary.}
\label{fig:13}
\end{figure}

\begin{figure} [htbp]
\begin{minipage}{\columnwidth}
\centering
\includegraphics[scale=0.33]{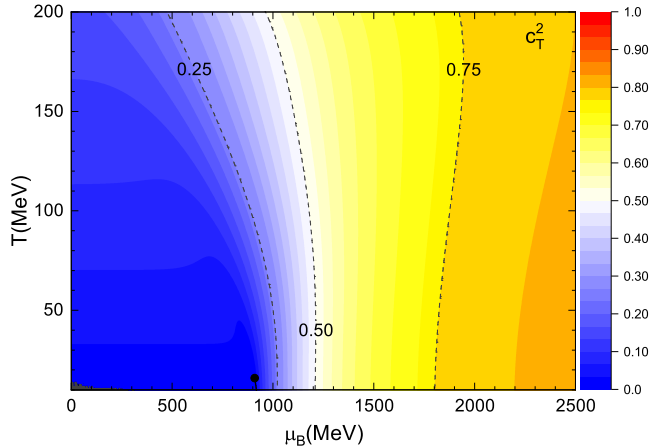}
\end{minipage}
\caption{ Contour map of $c^2_{T}$ in the  $T-\mu_B$ panel.}
\label{fig:14}
\end{figure}

Fig.~\ref{fig:13} and  Fig.~\ref{fig:14} demonstrate the contour maps of  $c^2_T$ in the $T-\rho_B$ and $T-\mu_B$ panel, respectively. The two figures show that the value of $c^2_T$ increases with the rising density or chemical potential in the stable phase, and the causality is always satisfied with  $c^2_T<1$.  The value  of  $c^2_T$ is close to   $c^2_{s/\rho_B}$ at  low temperature, since the isentropic trajectories are approximately  parallel to the density or chemical potential axis as indicated in Fig.~\ref{fig:1}.
The behavior of  $c^2_T$ in nuclear matter at low temperature is to some extent similar that of neutron star matter under $\beta$ equilibrium. Combined with the behavior of  $c^2_T$ in quark matter, if a hadron-quark phase transition happens at high density, the value of  $c^2_T$ will decrease in the mixed phase or pure quark phase. The detailed study about the speed of sound in neutron star matter is in progress.

Fig.~\ref{fig:13} also demonstrates that there exists a region~(gray area) where  $c^2_T=(\frac{\partial p}{\partial \epsilon})_T<0$. This region is just the unstable phase derived with the thermodynamic stability conditions. On the boundary of spinodal line including the CEP of the first-order transition, the value of $c^2_T$ is zero. $c^2_T$ is positive in both the stable and metastable phases.

\begin{figure} [htbp]
\begin{minipage}{\columnwidth}
\centering
\includegraphics[scale=0.33]{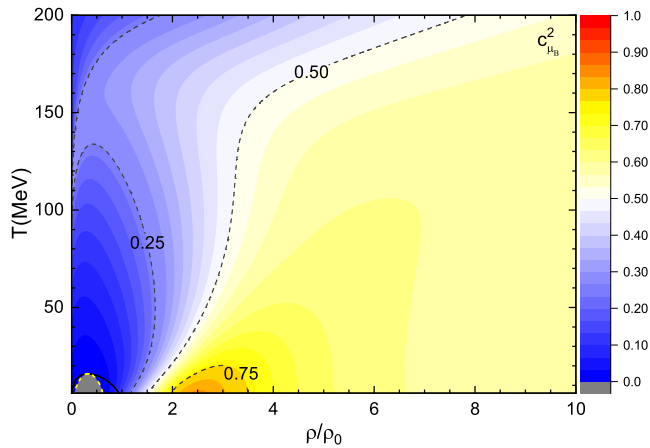}
\end{minipage}
\caption{ Contour map of $c^2_{\mu_B}$ in the  $T-\rho_B$ panel. The yellow dashed curve is the spinodal line and the boundary of $c^2_{\mu_B}=0$. $c^2_{\mu_B}=(\partial p / \partial \epsilon)_{\mu_B}$ is negative inside the boundary. }
\label{fig:15}
\end{figure}

\begin{figure} [htbp]
\begin{minipage}{\columnwidth}
\centering
\includegraphics[scale=0.33]{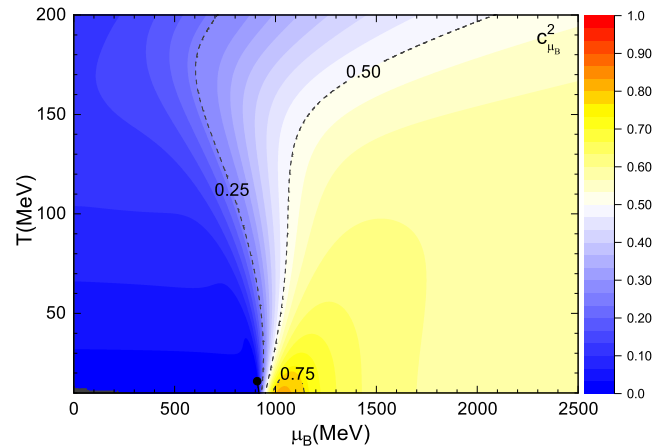}
\end{minipage}
\caption{ Contour map of $c^2_{\mu_B}$ in the  $T-\mu_B$ panel.}
\label{fig:16}
\end{figure}

We display the contour map of $c^2_{\mu_B}$ in Fig.~\ref{fig:15} and  Fig.~\ref{fig:16}. Similar to Fig.~\ref{fig:13}, the contour map in Fig.~\ref{fig:15} indicates that $c^2_{\mu_B}=(\partial p / \partial \epsilon)_{\mu_B}<0$ inside the spinodal line.  $c^2_{\mu_B}$ is positive in the metastable and stable phases. The contour maps in Fig.~\ref{fig:15} and  Fig.~\ref{fig:16} also indicate that the value of  $c^2_{\mu_B}$ has a peak-like structure in the phase diagram.

Finally, we note that the numerical results at high temperature should be treated with caution, since more hadronic degrees of freedom will appear at high temperature. In this study, we show the numerical results in the full phase diagram in order to demonstrate the whole changing trend of speed of sound. 

\section{Summary}   
In this work, we studied the speed of sound in symmetric nuclear matter at finite temperature and density~(chemical potential) in the nonlinear Walecka model.
We derived the speed of sound under different definitions in the complete phase diagram including the stable, metastable and unstable phases associated with the first-order phase transition.  We systematically discussed the relations between the speed of sound and nuclear LG phase transition.

The numerical results indicate that the behavior of the speed of sound in the phase diagram
is closely related the phase structure of nuclear matter. From the perspective of ideal fluid evolution, we focus on exploring the behavior of adiabatic sound velocity at constant $s/\rho_B$. The calculation indicates that the sound speed  $c^2_{s/\rho_B}$ is nonzero at the CEP under the mean field approximation, and the boundary of vanishing sound velocity is derived. We also found that the value of  $c^2_{s/\rho_B}$ is quite larger than 1/3 at high density, different from that in quark matter where  $c^2_{s/\rho_B}$ approaches to 1/3 at high density and/or high temperature.

We also explored the behaviors of sound speed under different physical conditions, and analyzed the correlations with the nuclear LG phase transition, as well as the relations between different definitions. The calculation shows that it is natural in nuclear matter to have a sound speed larger than $\sqrt{1/3}$ at a few times nuclear saturation density. A further study on the speed of sound in neutron star matter with a hadron-quark phase transition and observation constraints will be performed in the future.

\begin{acknowledgements} 
This work is supported by the National Natural Science Foundation of China under
Grant No. 11875213.
\end{acknowledgements}

\end{document}